\documentclass[final]{elsarticle}

\usepackage{natbib}
\biboptions{sort&compress}
\usepackage{graphicx}
\usepackage{subcaption}
\usepackage{amstext}
\usepackage{amssymb}
\usepackage{amsmath}
\usepackage{color}
\usepackage{booktabs}
\usepackage{todonotes}
\usepackage{siunitx}
\usepackage{mhchem}
\usepackage[displaymath]{lineno}
\usepackage{tikz}
\usetikzlibrary{math}

\presetkeys%
    {todonotes}%
    {inline}{}

\makeatletter
\renewcommand*\env@matrix[1][*\c@MaxMatrixCols c]{%
  \hskip -\arraycolsep
  \let\@ifnextchar\new@ifnextchar
  \array{#1}}
\makeatother

\newcommand{\porousblock}[3]{
  \definecolor{myblue}{RGB}{41,128,185}
  \definecolor{mygray}{RGB}{200,200,200}
  \pgfmathsetseed{#3};
  \coordinate (c) at (#1);
  \def\t{1.5};
  \def\rad{0.08};
  \fill[mygray] (c) --+ (0,-#2) --+ (-#2,-#2) --+ (-#2,0) -- (c);
  \fill[mygray] (c) --+ (0.25*\t,0.125*\t) --+  (-#2+0.25*\t,0.125*\t)
  --+ (-#2,0);
  \fill[mygray] (c)+(0,-#2) --+ (0.25*\t,-#2+0.125*\t)
  --+ (0.25*\t,0.125*\t) -- (c);   
  \draw[semithick] (c) --+ (0,-#2) --+ (-#2,-#2) --+ (-#2,0) -- (c);
  \draw[semithick] (c) --+ (0.25*\t,0.125*\t);
  \draw[semithick] (c)+(0.25*\t,0.125*\t) --+ (-#2+0.25*\t,0.125*\t);
  \draw[semithick] (c)+(-#2+0.25*\t,0.125*\t) --+ (-#2,0);
  \draw[semithick] (c)+(0,-#2) --+ (0.25*\t,-#2+0.125*\t)
  --+ (0.25*\t,0.125*\t);
  \pgfmathtruncatemacro{\nc}{3*#2}
  \foreach \xc in {1,2,...,\nc}{
    \foreach \yc in {1,2,...,\nc}{
      \pgfmathtruncatemacro{\rr}{5*rand}
      \ifthenelse{\rr > -2}{
        \draw[fill=white] (c) +
        (-4*\rad*\xc + rand*\rad + 1.5*\rad,
        -4*\rad*\yc + rand*\rad + 1.5*\rad)
        circle (\rad);}{
        \draw[fill=myblue] (c) +
        (-4*\rad*\xc + rand*\rad + 1.5*\rad,
        -4*\rad*\yc + rand*\rad + 1.5*\rad)
        circle (\rad);}
    }
  }
}

\renewcommand{\vec}[1]{\ensuremath{\mathbf{#1}}}

\newcommand{\w}{\ensuremath{\text{w}}}
\newcommand{\n}{\ensuremath{\text{n}}}
\newcommand{\p}{\ensuremath{\text{p}}}

\newcommand{\rel}{\ensuremath{\text{r}}}

\newcommand{\advective}{\ensuremath{\text{a}}}
\newcommand{\capillary}{\ensuremath{\text{c}}}
\newcommand{\ca}{\ensuremath{\text{Ca}}}

\newcommand{\dd}{\ensuremath{\text{d}}}
\newcommand{\D}{\ensuremath{\text{D}}}
\newcommand{\pd}[2]{\ensuremath{\frac{\partial #1}{\partial #2}}}

\newcommand{\od}[2]{\ensuremath{\frac{\dd #1}{\dd #2}}}

\journal{Transport in Porous Media}

\begin{document}


\begin{frontmatter}

\title{Pore network modeling of the effects of viscosity ratio and
  pressure gradient on steady-state incompressible two-phase flow in
  porous media}

\author[ntnu-ify]{Magnus Aa.\ Gjennestad\corref{mag-corresp}}
\ead{magnus@aashammer.net}
\author[ntnu-ify]{Mathias Winkler}
\author[ntnu-ify]{Alex Hansen}

\cortext[mag-corresp]{Corresponding author.}

\address[ntnu-ify]{PoreLab and Department of Physics, Norwegian
  University of Science and Technology, \\ Trondheim, Norway}

\begin{abstract}
  We perform more than 6000 steady-state simulations with a dynamic
  pore network model, corresponding to a large span in viscosity
  ratios and capillary numbers. From these simulations, dimensionless
  quantities such as relative permeabilities, residual saturations,
  mobility ratios and fractional flows are computed. Relative
  permeabilities and residual saturations show many of the same
  qualitative features observed in other experimental and modeling
  studies. However, while other studies find that relative
  permeabilities converge to straight lines at high capillary numbers
  we find that this is not the case when viscosity ratios are
  different from 1. Our conclusion is that departure from straight
  lines occurs when fluids mix rather than form decoupled flow
  channels. Another consequence of the mixing is that computed
  fractional flow curves, plotted against saturation, lie closer to
  the diagonal than they would otherwise do. At lower capillary
  numbers, fractional flow curves have a classical S-shape. Ratios of
  average mobility to their high-capillary number limit values are
  also considered. These vary, roughly, between $0$ and $1$, although
  values larger than $1$ are also observed. For a given saturation and
  viscosity ratio, the mobilities are not always monotonically
  increasing with the pressure gradient. While increasing the pressure
  gradient mobilizes more fluid and activates more flow paths, when
  the mobilized fluid is more viscous, a reduction in average mobility
  may occur.
\end{abstract}


\end{frontmatter}

\section{Introduction}
A number of different modeling approaches have been applied to study
two-phase flow in porous media. These include direct numerical
simulations (DNS), which employ e.g.\ the volume-of-fluid method
\citep{Raeini2012} or the level-set method
\citep{Jettestuen2013,Gjennestad2015} to keep track of the fluid
interfaces, lattice-Boltzmann methods \citep{Ramstad2012} and pore
network models. Recently, a number methods were compared in a
benchmark study by \citet{Zhao2019}, where participants were asked to
reproduce experimentally studied transient fluid displacement
processes at different capillary numbers and wettability conditions,
i.e.\ contact angles. The conclusion was that no single method was
successful under all conditions and that thin films and corner flow
posed substantial computational and modeling challenges.

This benchmark study, and the bulk of works in the literature, focus
on transient processes. Less attention has been given to pore-scale
modeling and experiments in steady-state flow, i.e.\ flow where
macroscopic quantities such as fractional flow fluctuate around a
well-defined mean. On the modeling side, part of the explanation is
probably that steady-state simulations require large systems and
longer simulation times compared to transient processes. While
break-through of the invading phase typically happens for simulation
times corresponding to much less than one pore volumes of flow in
transient cases, several pore volumes may be required to obtain decent
time-averages of steady-state quantities.

In spite of this, some studies on steady-state two-phase flow have
been done. \citet{Avraam1995} did quasi-2D micro model experiments,
varied the capillary number, the viscosity ratio and the flow rate
ratio, and found four different flow regimes. They also studied
relative permeabilities. Steady-state simulations with a pore network
model of the Aker type \citep{Aker1998} have also been performed by
e.g.\ \citet{Knudsen2002}, \citet{Knudsen2002b} and
\citet{Ramstad2006}. In particular, \citet{Knudsen2002} did
simulations with equal viscosities and one value for the interfacial
tension, and studied effect of changing total flow rate on
e.g.\ fractional flow and relative permeabilities. Results for equal
viscosities are interesting and applicable in some cases, e.g.\ for
mineral oil and water \citep{Oak1990}. In other applications,
e.g.\ sequestration of supercritical \ce{CO2} \citep{Bennion2005} and
gas-liquid flows such as in fuel cells, they are not.

We present results from more than 6000 steady-state simulations, that
cover a large range of viscosity ratios and capillary numbers. The
chosen pore network model is also of the Aker type \citep{Aker1998},
specifically the variant described by \citet{Gjennestad2018b}. Other
variants of the Aker model can be found in
\citep{Sinha2019,Knudsen2002,Knudsen2002b,Ramstad2006}. The model has
several properties that are advantageous when computing steady-state
quantities. First, it is dynamic and thus captures the effects of both
viscous and capillary forces. Second, it can be solved in a
numerically stable manner at arbitrarily low capillary numbers
\citep{Gjennestad2018b}. Third, it is possible to apply periodic
boundary conditions, keeping the saturation constant and eliminating
effects of saturation gradients. Furthermore, it is computationally
cheap, making the study of large enough systems over long enough times
possible.

In spite of these advantages, however, the model also has some
limitations. In particular, film flow is not accounted for and the
construction of the model makes it difficult to capture accurately
cooperative pore-filling events during imbibition
\citep{Zhao2019}. While film flow effects could, in principle, be
captured e.g.\ by a DNS or lattice-Boltzmann method, very high spatial
resolution is required to resolve such films properly
\citep{Zhao2019}. This makes such an approach prohibitively expensive
for steady-state calculations, especially when a large number of them
are desired. Film flow could, in principle, also be included in the
present model \citep{Tora2012}. However, use of this modified model at
low capillary numbers would probably require the construction of a new
solution method to ensure numerical stability. Cooperative pore
filling events were more accurately captured by other models in the
benchmark study \citep{Zhao2019}. However, these relied on quasi-static
considerations, making them difficult to apply directly in a
steady-state simulation.

In the simulations, we utilize a recent innovation in the numerical
solution method \citep{Gjennestad2018b} to perform numerically stable
simulations at low and moderate capillary numbers. The new methodology
has an important effect at capillary numbers below \SI{e-3}{}. In
addition, we make extensive use of a recent study of the
high-capillary number regime \citep{Sinha2018} in the analysis of the
results. The discussion is restricted to capillary numbers above
\SI{e-4}{}, where history-dependence of the steady-state quantities is
negligible \citep{Knudsen2002}. At lower capillary numbers,
steady-state quantities are harder to define and calculate. To allow
for a discussion which is as general as possible, and which allows for
comparison with other studies of slightly different systems, we focus
on dimensionless steady-state quantities, such as relative
permeabilities, mobility ratios and fractional flow\footnote{These
  quantities are used to provide a familiar framework of dimensionless
  quantities in which results are presented and discussed. However,
  other quantities could also, in principle, be used to convey the
  same information. One example is the velocities presented
  in~\citep{Hansen2018}.}. To aid further research, the simulation data
are published along with this article.

The aim is to shed light on how different steady-state flow properties
behave as capillary numbers are changed from moderate values around
\SI{e-3}{}-\SI{e-4}{} to the high capillary number limit and to assess
the impact of viscosity ratio in this context. One important finding
is that relative permeabilities are not necessarily straight lines at
high capillary numbers. Our conclusion is that this occurs when fluids
have different viscosities and exhibit some degree of mixing rather
than forming separate flow channels. Another interesting finding is
that the average mobility, for a given saturation and viscosity ratio,
is not always a monotonically increasing function of the pressure
gradient. Intuitively, one might think this should be the case, as
increasing the pressure gradient mobilizes more fluid and activates
more flow paths. However, when the mobilized fluid is more viscous, a
reduction in average mobility may occur instead.

The rest of the paper is structured as follows. In
Section~\ref{sec:system}, we describe the system under consideration
and define some important steady-state flow properties. In
Section~\ref{sec:pnm}, we briefly describe the pore network model used
and the numerical methods used to solve it. The performed simulations
are described in Section~\ref{sec:simulations}. Results are presented
and discussed in Section~\ref{sec:results} and concluding remarks are
given in Section~\ref{sec:conclusion}.

\section{System}
\label{sec:system}

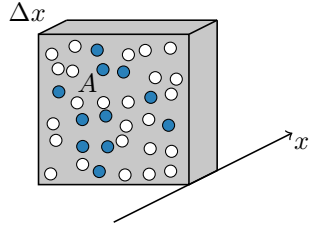
\begin{figure}[tbp]
  \centering
  \begin{tikzpicture}
    \porousblock{0,0}{2.0}{1337}
    \node (L) at (-2.15, 0.3) {$\Delta x$};
    \node (A) at (-1.35, -0.65) {$A$};
    \node (x) at (1.5, -2+0.5 + 0.25) {};
    \node[below] at (x) {$x$};
    \draw[semithick,->] (-1,-2-0.75 + 0.25) -- (x);
  \end{tikzpicture}
  \caption{Illustration of the system under consideration, a block of
    porous material. The porous matrix is shown in gray, pores filled
    with the wetting fluid in white and pores filled with the
    non-wetting fluid in blue. The block has thickness $\Delta x$ in
    the $x$-direction and cross sectional area $A$.}
  \label{fig:rev}
\end{figure}

The system we consider is a block of porous material, as illustrated
in Figure~\ref{fig:rev}. It has cross sectional area $A$ and thickness
$\Delta x$ in the direction of flow (the $x$-direction). The volume of
the block is
\begin{linenomath} \begin{align}
  \label{eq:V}
  V &= A \Delta x.
\end{align} \end{linenomath} 
The pore space volume in the block is $V_\p$, so that the porosity is
\begin{linenomath} \begin{align}
  \label{eq:phi}
  \varphi &= V_\p/V.
\end{align} \end{linenomath} 
The pore space is filled with two fluids, where one is more wetting
towards the pore walls than the other. In the following, we will call
the more wetting fluid wetting ($\w$) and the less wetting fluid
non-wetting ($\n$). The fluids are assumed to be incompressible and
$S_\w$ is the wetting fluid saturation, i.e.\ the fraction of the pore
space volume occupied by the wetting fluid.

A pressure difference $\Delta p$, either constant or fluctuating,
exists across the porous block. This causes the wetting and
non-wetting fluids to flow at rates $Q_\w$ and $Q_\n$,
respectively. The total flow rate is,
\begin{linenomath} \begin{align}
  Q = Q_\w + Q_\n,
\end{align} \end{linenomath} 
and the fractional flow of wetting fluid is
\begin{linenomath} \begin{align}
  F_\w = Q_\w / Q.
\end{align} \end{linenomath} 
The average fluid velocity in the pore space, the seepage velocity, is
\begin{linenomath} \begin{align}
  v &= Q / \varphi A.
\end{align} \end{linenomath}

\section{Pore network model}
\label{sec:pnm}

In this section, we briefly describe the pore network model used in
this study. For a more detailed description of the model and the
numerical methods used to solve it, the reader is referred
to~\citep{Gjennestad2018b}. An in-depth discussion of a slightly
different model, which is also of the Aker type~\citep{Aker1998}, can
be found in~\citep{Sinha2019}. Both models were recently used to study
the high capillary number regime~\citep{Sinha2018}.

\begin{figure}[tbp]
  \centering
  \begin{subfigure}[b]{0.48\textwidth}
    \includegraphics[width=\textwidth]{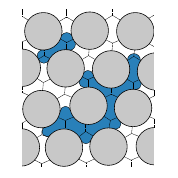}
    \caption{}
  \end{subfigure}
  ~
  \begin{subfigure}[b]{0.48\textwidth}
    \includegraphics[width=\textwidth]{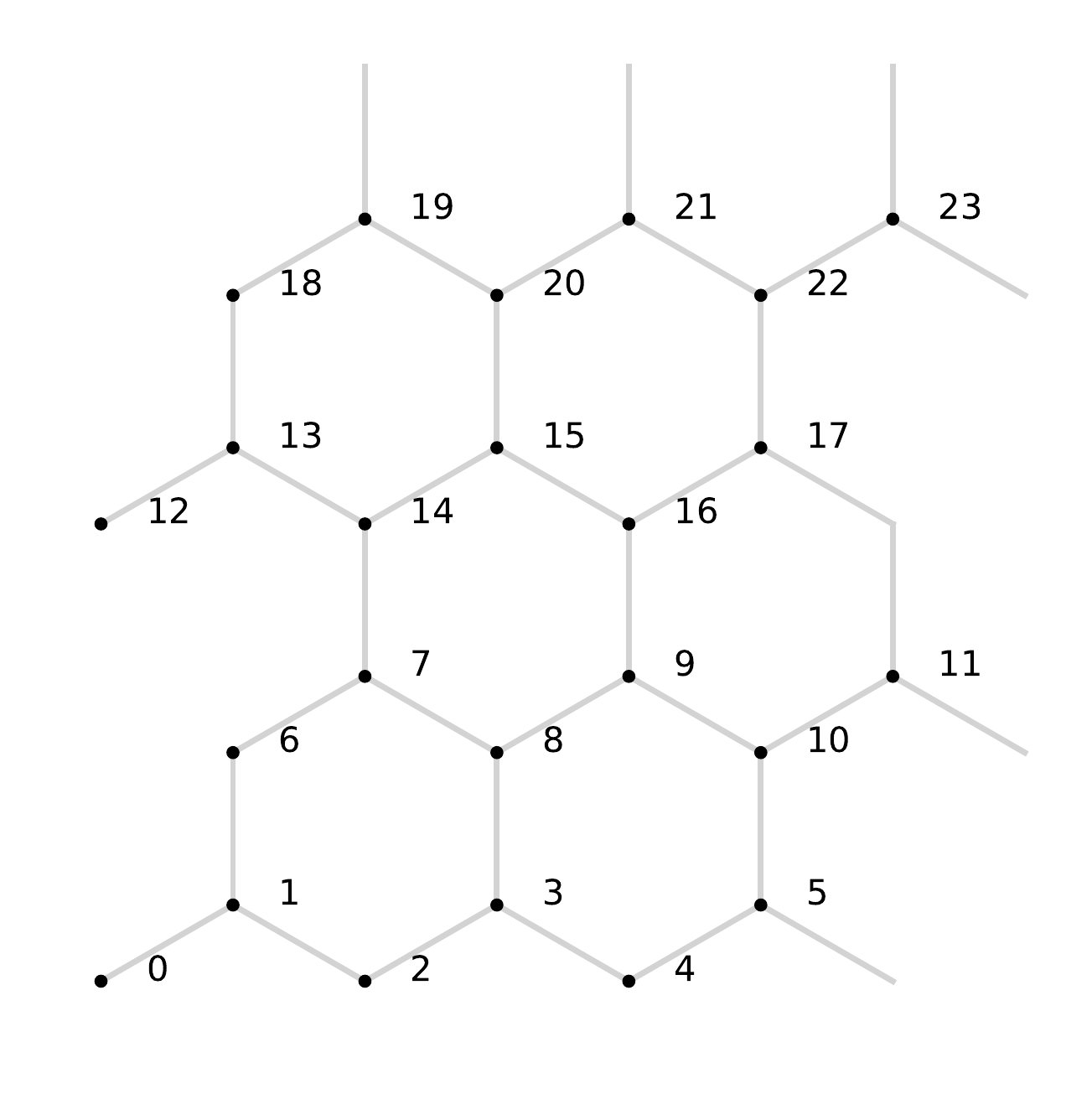}
    \caption{}
    \label{fig:network_indexing}
  \end{subfigure}
  \caption{Illustration of (a) wetting (white) and non-wetting fluid
    (blue) in a physical pore network and (b) the representation of
    this network in the model. The dashed lines in (a) indicate
    sections of the pore space volume that are each represented by one
    link in (b). The intersection points of the dashed lines in (a)
    show the node locations in the model representation (b). Figures
    (a) and (b) are reproduced from \citep{Gjennestad2018b}.}
  \label{fig:network_model}
\end{figure}

The model describes flow of two incompressible and immiscible fluids
($\w$ and $\n$) in a porous medium. The porous medium is represented
by a network consisting of $N$ nodes that are connected by $M$
links. The nodes are each given an index $i \in \left[0, ...,
  N-1\right]$. The links are identified by the two nodes $ij$ that
they connect. An example pore network is shown in
Figure~\ref{fig:network_model}. The nodes have no volume, and the pore
space volume is thus assigned to the links. It is assumed that each
fluid fills the entire link cross sections. The location of a
fluid-fluid interface can therefore be described by a single number
which gives its position in the link. For each link, the vector
$\vec{z}_{ij}$ contains the positions of the fluid interfaces in that
link.

The flow in the links is treated in a one-dimensional fashion,
averaged over the link cross sections. We consider flows in relatively
small cross sections only and therefore neglect any effects of fluid
inertia. The volumetric flow rate from node $j$ to node $i$ through
the link connecting the two nodes is then given by,
\begin{linenomath} \begin{align}
  \label{eq:pnm_q_ij}
  q_{ij} &= -\lambda_{ij} \left( \vec{z}_{ij} \right) \left\{ p_i - p_j -
  c_{ij} \left( \vec{z}_{ij} \right) \right\}.
\end{align} \end{linenomath} 
Herein, $p_i$ is the pressure in node $i$, $\lambda_{ij}$ is the
link's mobility and $c_{ij}$ is the net pressure difference across the
link due to its fluid interfaces. Both $\lambda_{ij}$ and $c_{ij}$
depend on the interface positions $\vec{z}_{ij}$. For two nodes $i$
and $j$ not connected by a link, $g_{ij} = 0$. Applying mass
conservation at each node $i$ yields,
\begin{linenomath} \begin{align}
  \label{eq:pnm_q_cons}
  \sum_j q_{ij} &= 0.
\end{align} \end{linenomath} 

The cross sectional area of link $ij$ is $a_{ij}$. The interface
positions $\vec{z}_{ij}$ therefore evolve in time according to the
advection equation,
\begin{linenomath} \begin{align}
  \label{eq:pnm_ode}
  \od{}{t} \vec{z}_{ij} = \frac{q_{ij}}{a_{ij}},
\end{align} \end{linenomath} 
when sufficiently far away from the nodes. Close to the nodes, they
are subject to additional models that account for interface
interactions in the nodes. This is described in
\citep{Gjennestad2018b}.

\subsection{Link mobility model}

The link mobility depends on link geometry and fluid viscosities. We
assume cylindrical links when computing the mobilities and thus
\begin{linenomath} \begin{align}
  \lambda_{ij} \left( \vec{z}_{ij} \right) &= \frac{\pi r_{ij}^4}{8 L_{ij}
    \mu_{ij} \left( \vec{z}_{ij} \right)}.
\end{align} \end{linenomath} 
Here, $L_{ij}$ is the link length, $r_{ij}$ is the link radius and
$\mu_{ij} \left( \vec{z}_{ij} \right)$ is the volume-weighted average
of the fluid viscosities $\mu_\w$ and $\mu_\n$.

\subsection{Interfacial pressure discontinuity model}

There may be zero, one or more interfaces in each link. Their
positions along the link are contained in $\vec{z}_{ij}$. Each element
in $\vec{z}_{ij}$ is thus between $0$ and $L_{ij}$. The symbol
$c_{ij}$ denotes the sum of the interfacial pressure discontinuities
in link $ij$. We assume that the links are much wider near the ends
than in the middle and that the pressure discontinuities become
negligibly small for interfaces near the ends. The pressure
discontinuities are therefore modeled by
\begin{linenomath} \begin{align}
  c_{ij} \left( \vec{z}_{ij} \right) &= \frac{2 \sigma_{\w\n}}{r_{ij}}
  \sum_{z \in \vec{z}_{ij}} \left( \pm 1 \right) \left\{ 1 - \cos
  \left( 2 \pi \chi \left( z \right) \right) \right\}.
\end{align} \end{linenomath} 
Herein, $\sigma_{\w\n}$ is the interfacial tension and
\begin{linenomath} \begin{align}
  \chi \left( z \right) &=
  \begin{cases}
    0, & \text{if} \ z < \beta r_{ij}, \\ \frac{z - \beta
      r_{ij}}{L_{ij} - 2 \beta r_{ij}}, & \text{if} \ \beta r_{ij} <
    z < L_{ij} - \beta r_{ij}, \\ 1, & \text{if} \ z > L_{ij} -
    \beta r_{ij}.
  \end{cases}
\end{align} \end{linenomath} 
The effect of the $\chi$-function is to introduce zones of length
$\beta r_{ij}$ at each end of the links where the pressure
discontinuity of any interface is zero.

\subsection{Boundary conditions}
\label{sec:boundary_conditions}

In this study, we will run steady-state simulations in a network that
can be laid out in two dimensions, as illustrated in
Figure~\ref{fig:network_indexing}. The network is periodic both in the
flow direction and in the transverse direction. A pressure difference
of $\Delta p$ will be applied across the periodic boundary in the flow
direction, or a total flow rate $Q$ will be prescribed, as described
in \citep{Gjennestad2018b}. The length of the network in the flow
direction is denoted $\Delta x$ and the average pressure gradient in
the network is thus $\Delta p/\Delta x$.

\subsection{Numerical solution method}

Inserting \eqref{eq:pnm_q_ij} into \eqref{eq:pnm_q_cons} gives a
system of equations for the unknown node pressures. The exact form of
this system depends on the numerical method to be used. Here, we will
use the Forward Euler method, where the length of time step $n$ is set
according to the criterions derived in \citep{Gjennestad2018b},
\begin{linenomath}
  \begin{align}
    \Delta t^{(n)} = \min \left( \Delta t^{(n)}_\capillary, \Delta
    t^{(n)}_\advective \right).
  \end{align}
\end{linenomath}
Herein,
\begin{linenomath}
  \begin{align}
    \label{eq:dt_a}
    \Delta t^{(n)}_\advective &= C_\advective \min_{ij} \left(
    \frac{a_{ij} L_{ij}}{q_{ij}^{(n)}} \right), \\
    \label{eq:dt_c}
    \Delta t^{(n)}_\capillary &= C_\capillary \min_{ij} \left( \frac{2
      a_{ij} }{ \lambda_{ij}^{(n)} \left| \sum_{z \in
        \vec{z}_{ij}^{(n)}} \pd{c_{ij}}{z} \right|} \right),
  \end{align}
\end{linenomath}
and the parameters $C_\advective$ and $C_\capillary$ are set to $0.1$
and $0.9$, respectively, which together ensure numerical
stability. Once the system is solved and the node pressures are
obtained, the link flow rates can be calculated from
\eqref{eq:pnm_q_ij} and the fluid interfaces moved according to
\eqref{eq:pnm_ode}. Further details can be found
in~\citep{Gjennestad2018b}.

\subsection{Computation of average quantities from network simulations}
\label{sec:pnm_measurements}
The porous medium we consider is a network of links, and the total
volume of the links is the pore volume $V_\p$. The network is embedded
in a three dimensional block of solid material with thickness $\Delta
x$ in the flow direction and cross sectional area $A$. The volume $V$
of the porous block and its porosity $\varphi$ are then easily
calculated by \eqref{eq:V} and \eqref{eq:phi}, respectively.

The saturation $S_\w$ may be computed at any time during the
simulation, by adding up the fluid volumes for all links. However,
since we here use periodic boundary conditions, $S_\w$ is a constant
in each simulation. So is $S_\n = 1 - S_\w$.

In the case of constant applied pressure gradient $\Delta p/\Delta x$,
the quantities that we need to compute from the actual simulations are
$Q$, $Q_\w$ and $Q_\n$. These are time-averages of fluctuating
quantities. The model is stepped forward in time as described
in the previous section. We calculate the time-average $Q$ by
summing over the total flow rates $Q^{\left(n\right)}$ at each time
step $n$ (after steady-sate has been reached),
\begin{linenomath} \begin{align}
  Q = \frac{\sum_n Q^{\left(n\right)} \Delta t^{\left(n\right)}
  }{\sum_n \Delta t^{\left(n\right)}}.
\end{align} \end{linenomath} 
The time-averaged quantities $Q_\w$ and $Q_\n$ is calculated from
$Q_\w^{\left(n\right)}$ and $Q_\w^{\left(n\right)}$ in an analogous
manner.

The instantaneous flow rate $Q^{\left(n\right)}$ can be computed by
constructing a plane cutting through the network, transverse to the
flow direction, and adding together the flow rates
$q_{ij}^{\left(n\right)}$ of all links intersecting the plane. We
denote the set of intersecting links by $B$ and add up,
\begin{linenomath} \begin{align}
  Q^{\left(n\right)} = \sum_{ij \in B} q_{ij}^{\left(n\right)}.
\end{align} \end{linenomath} 
Since the fluids are incompressible, it does not matter where this cut
is made.

The instantaneous flow rate $Q^{\left(n\right)}_\w$ is computed by
making several cuts, denote the set of cuts by $C$, and computing the
sum
\begin{linenomath} \begin{align}
  Q^{\left(n\right)}_\w = \frac{1}{\left| C \right|} \sum_{B \in C}
  \sum_{ij \in B} s_{ij}^{\left(n\right)} q_{ij }^{\left(n\right)}.
\end{align} \end{linenomath} 
Herein, $\left| C \right|$ denotes the number of elements in $C$,
i.e.\ the number of cuts, and $s_{ij}^{\left(n\right)}$ is the volume
fraction of wetting fluid in the volume of fluid that flowed past the
middle of link $ij$ during time step $n$. $Q_\n^{\left(n\right)}$ is
computed in an analogous manner. Having computed the time-averages
$Q$, $Q_\w$ and $Q_\n$ we may the obtain the time-averaged flow
velocity, mobility, fractional flow and relative permeabilities.

If $Q$ is fixed instead of $\Delta p/\Delta x$, the time-averaged
value of the pressure gradient is computed by
\begin{linenomath} \begin{align}
  \frac{\Delta p}{\Delta x} = \frac{\sum_n \Delta p^{\left(n\right)}
    \Delta t^{\left(n\right)} }{\Delta x \sum_n \Delta
    t^{\left(n\right)}},
\end{align} \end{linenomath} 
where $\Delta p^{\left(n\right)}$ is the pressure difference across
the network during time step $n$.

Using average quantities calculated as described above, the capillary
number is computed according to,
\begin{linenomath} \begin{align}
  \ca &= \frac{\bar{\mu} \left| Q \right|}{\varphi A \sigma_{\w\n}},
\end{align} \end{linenomath} 
where the average viscosity is defined as,
\begin{linenomath} \begin{align}
  \bar{\mu} &= S_\w \mu_\w + S_\n \mu_\n.
\end{align} \end{linenomath} 

\subsection{Dimensional analysis}
\label{sec:dimensional_analysis}

As can be surmised from the description above, the network and five
numbers are given as input to steady-state simulations. In the case of
constant pressure-difference boundary conditions, the five numbers are
the fluid viscosities $\mu_\w$ and $\mu_\n$, the fluid-fluid
interfacial tension $\sigma_{\w\n}$, the pressure gradient $\Delta
p/\Delta x$ and the saturation $S_\w$. Any change in the steady-state
averages is the response of the model to variations in these
inputs. If we consider the network topology and aspect ratios fixed,
and only allow for a linear scaling of the network size, any
variations in the network can be described by a single length
scale. We here choose the average pore radius $\bar{r}$.

By the Buckingham $\pi$ theorem \citep{Rayleigh1892}, the total of six
dimensional input variables can be reduced to three dimensionless
variables. This means that any combination of the six inputs that give
the same three dimensionless variables are similar and differ only in
scale. Any dimensionless output from the model is therefore the same
for the same set of dimensionless input variables. One choice of
dimensionless variables is
\begin{linenomath} \begin{align}
  S_\w, & \\ M &= \frac{\mu_\n}{\mu_\w}, \\ \Pi &= \left| \frac{\Delta
    p}{\Delta x} \right| \frac{\bar{r}^2}{2 \sigma_{\w\n}},
\end{align} \end{linenomath} 
where $M$ is the viscosity ratio. The variable $\Pi$ is a dimensionless
pressure gradient. It represents the ratio of the average pressure
drop over a length $\bar{r}$ to the Young--Laplace pressure difference
over an interface in a pore of radius $\bar{r}$. In particular, when
$\Pi = 1$, we have
\begin{linenomath} \begin{align}
  \left| \frac{\Delta p}{\Delta x} \right| \bar{r} = \frac{2
    \sigma_{\w\n}}{\bar{r}},
\end{align} \end{linenomath} 
and the average pressure drop over the length $\bar{r}$ is equal to
the typical Young--Laplace pressure difference.

Since it relates the average pressure drop to the capillary forces,
$\Pi$ may be expected to play a similar role as the capillary
number. This should be true at least when capillary numbers are high
and the average pressure drop is dominated by viscous
contributions. However, $\Pi$ is perhaps more closely related to the
ganglion mobilization number. This was defined by \citet{Avraam1995}
as the ratio between the driving force exerted on a ganglion and its
resistance to motion resulting from capillary forces.

\section{Simulations}
\label{sec:simulations}

Steady-state simulations were performed using the pore network model
described in Section~\ref{sec:pnm}. All simulations were run on $72
\times 48$ hexagonal networks, similar to that shown in
Figure~\ref{fig:network_indexing}. These networks consisted of 3456
nodes and 5184 links. All links had the same length $L$ and link radii
were uniformly distributed between $0.1L$ and $0.4L$. In total, 6048
simulations were run with input parameters in the ranges given in
Table~\ref{tbl:pnm_parameters}. For each of the 288 combinations of
the input parameters, 21 values of $S_\w$ were used, evenly spaced on
the interval $\left[0, 1\right]$. Time-averaged quantities were
calculated from simulation results as described in
Section~\ref{sec:pnm_measurements}. The averaging time corresponded to
$10$ pore volumes of flow.

\begin{table}[tbp]
  \center
  \caption{Range of input parameters used in the steady-state pore
    network model simulations. For each combination of the input
    parameters, 21 values of $S_\w$, evenly spaced on the interval
    $\left[0, 1\right]$, were used. The corresponding ranges of the
    dimensionless variables $M$, $\Pi$ and $\ca$ are also given (below
    the horizontal line).}
  \label{tbl:pnm_parameters}
  \begin{tabular}{l l l l}
    \toprule
    Quantity & Minimum value & Maximum value & Unit \\
    \midrule
    $\mu_\w$ & \SI{5.0e-4}{} & \SI{1.0e-2}{} & \si{\pascal\second} \\
    $\mu_\n$ & \SI{5.0e-4}{} & \SI{1.0e-2}{} & \si{\pascal\second} \\
    $\sigma_{\w\n}$ & \SI{2.0e-2}{} & \SI{3.0e-2}{} & \si{\newton\per\meter} \\
    $-\Delta p/\Delta x$ & \SI{3.9e3}{} & \SI{8.0e5}{} &
    \si{\pascal\per\meter} \\
    $\bar{r}$ & \SI{2.5e-4}{} & \SI{7.8e-4}{} & \si{\meter} \\
    \midrule
    $M$ & \SI{5.0e-2}{} & \SI{2.0e1}{} & - \\
    $\Pi$ & \SI{6.1e-3}{} & \SI{8.1e1}{} & - \\
    $\ca$ & \SI{4.0e-4}{} & \SI{6.1e-1}{} & - \\
    \bottomrule
  \end{tabular}
\end{table}

\section{Results}
\label{sec:results}

In this section, we present and discuss the simulation results. We
look first at relative permeabilities
(Section~\ref{sec:relative_permeabilities}), then residual saturations
(Section~\ref{sec:residual_saturations}) average flow velocities and
mobilities (Section~\ref{sec:mobilities}) and, finally, fractional
flows (Section~\ref{sec:fractional_flows}).

\subsection{Relative permeabilities}
\label{sec:relative_permeabilities}

\begin{figure}[tbp]
  \centering
  \begin{subfigure}[b]{0.48\textwidth}
    \includegraphics[width=\textwidth]{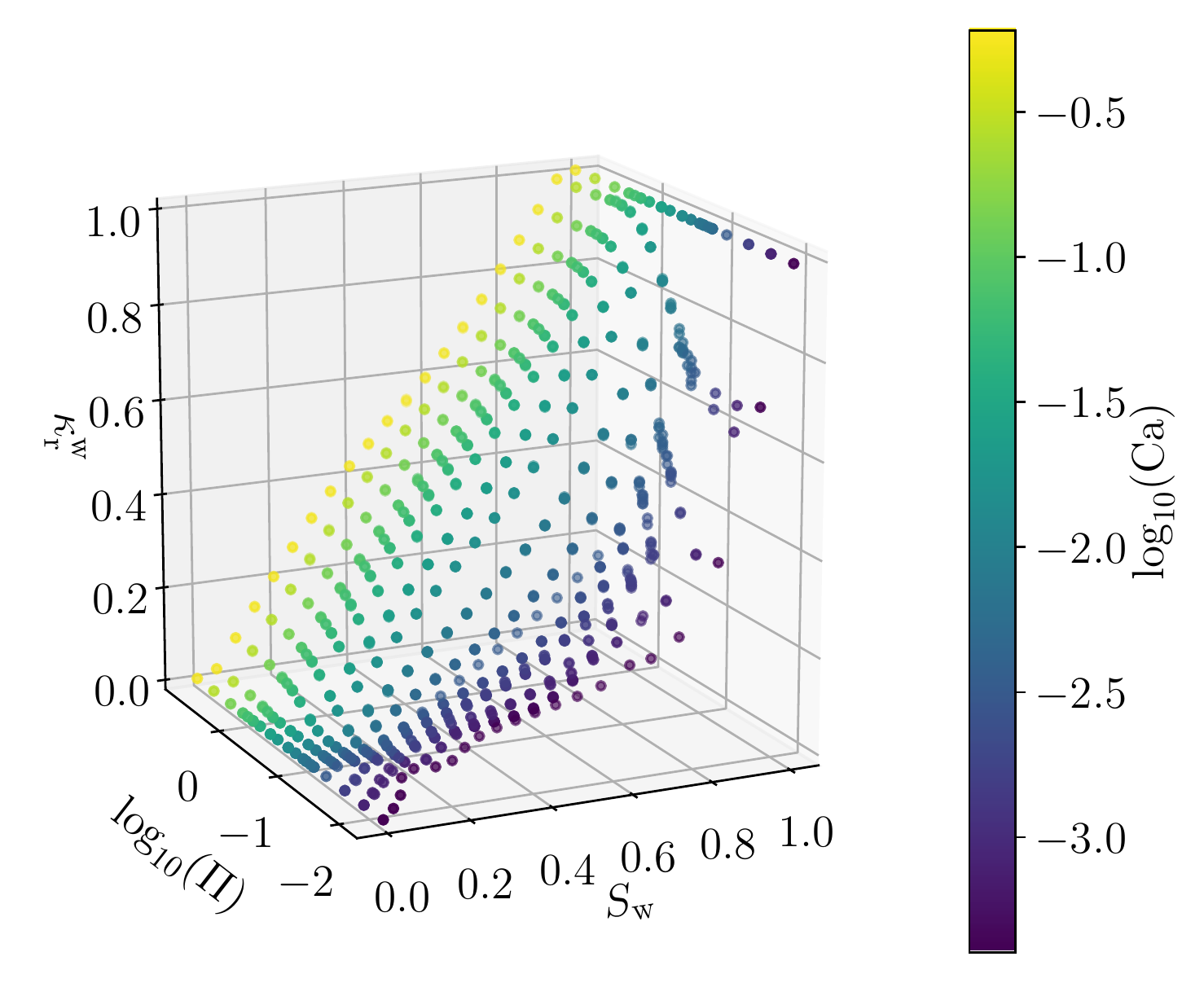}
    \caption{}
    \label{fig:kr_w_M_1_pnm}
  \end{subfigure}
  \begin{subfigure}[b]{0.48\textwidth}
    \includegraphics[width=\textwidth]{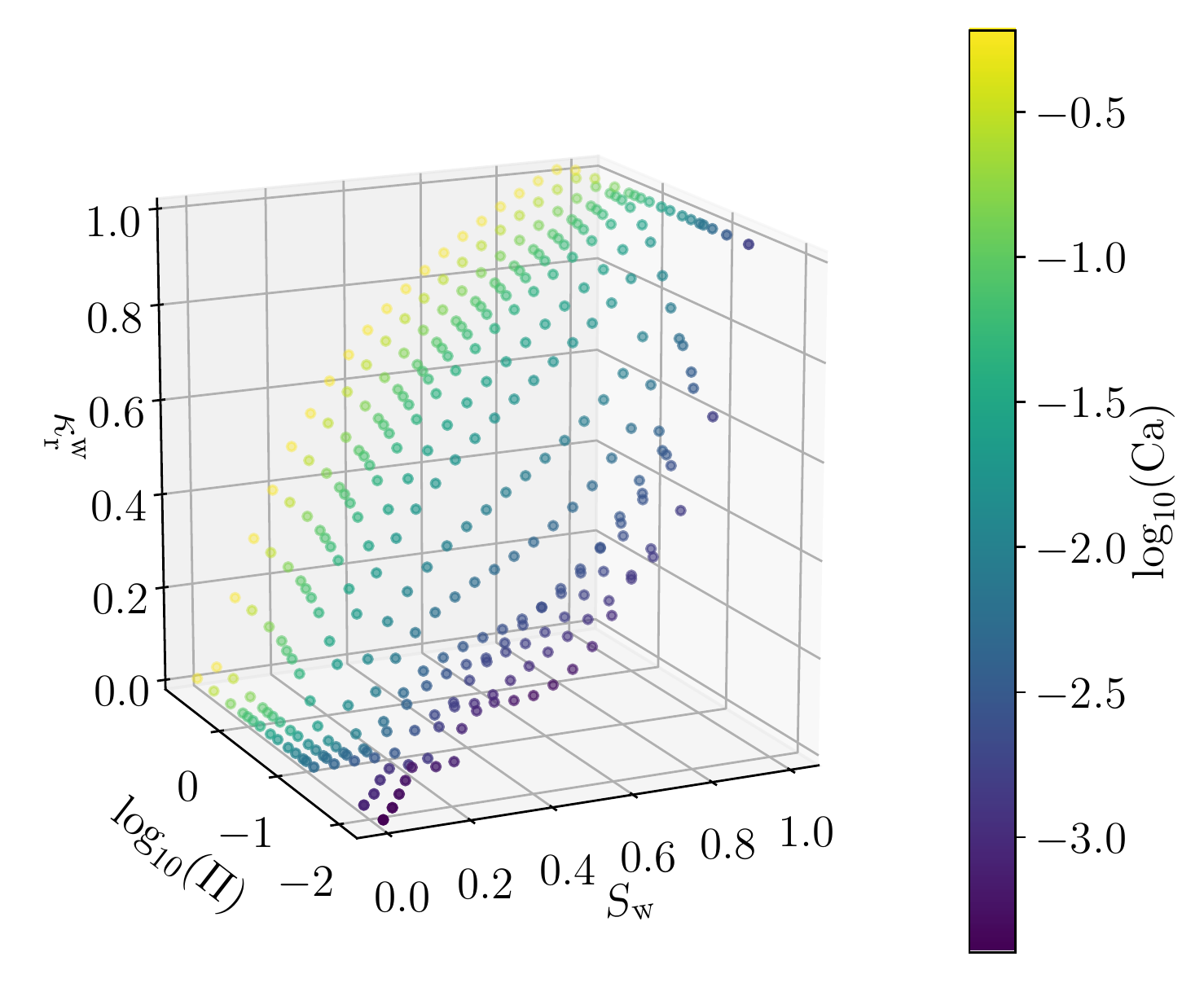}
    \caption{}
    \label{fig:kr_w_M_0_25_pnm}
  \end{subfigure}
  \begin{subfigure}[b]{0.48\textwidth}
    \includegraphics[width=\textwidth]{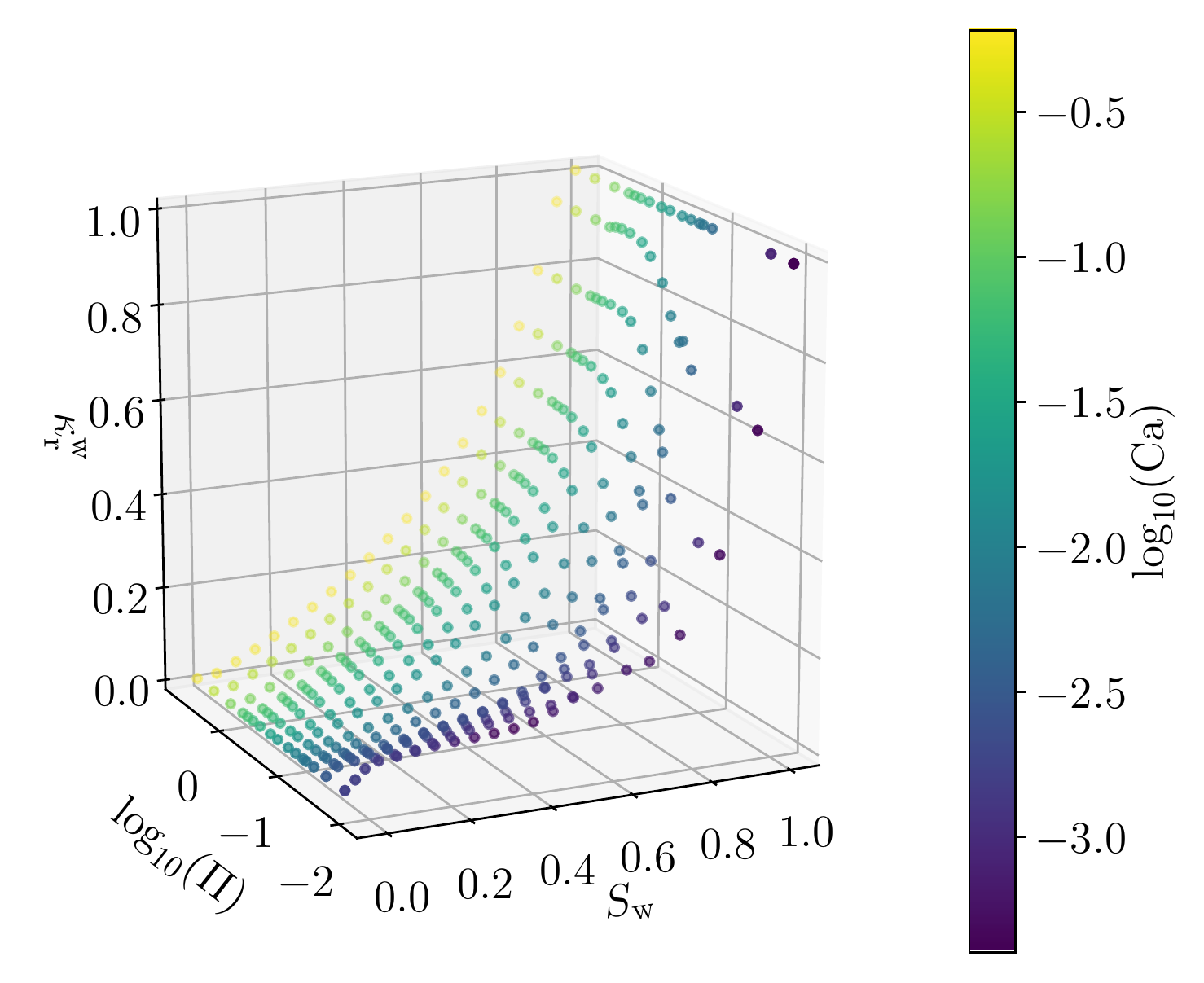}
    \caption{}
    \label{fig:kr_w_M_4_pnm}
  \end{subfigure}
  \begin{subfigure}[b]{0.48\textwidth}
    \includegraphics[width=\textwidth]{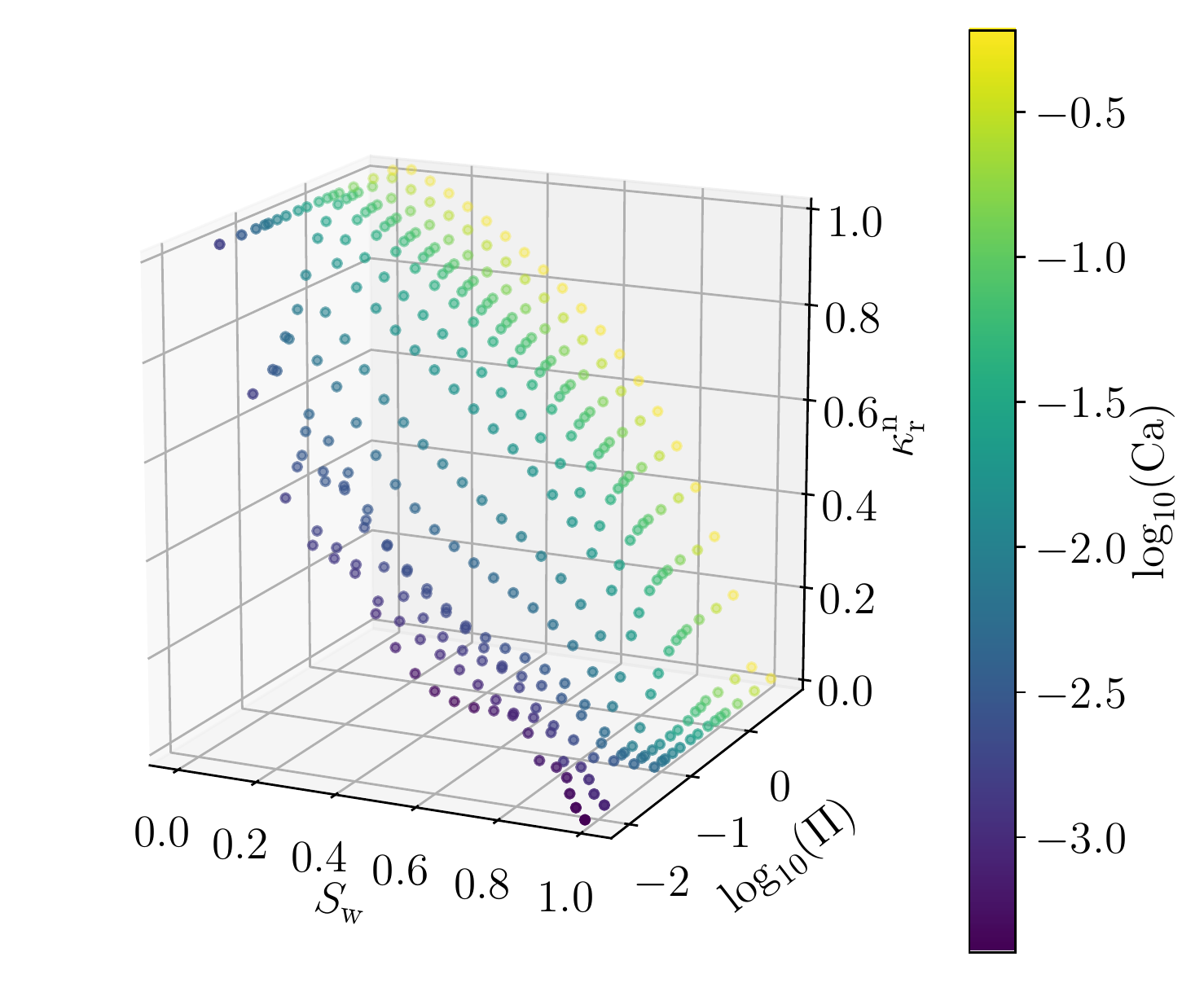}
    \caption{}
    \label{fig:kr_n_M_4_pnm}
  \end{subfigure}
  \caption{Relative permeabilities for the wetting phase for (a)
    $M=1$, (b) $M=0.25$ and (c) $M=4$. Relative permeabilities for the
    non-wetting phase and $M=4$ are shown in (d).}
  \label{fig:kr}
\end{figure}

Relative permeabilities $\kappa^\rel_\w$ and $\kappa^\rel_\n$ are
perhaps the most extensively studied properties in two-phase flow in
porous media, and the most obvious dimensionless numbers to calculate
from the pore network model. They relate the flow rates of each phase
to the pressure drop through
\begin{linenomath} \begin{align}
  \frac{Q_\w}{A} &= - \frac{\kappa^\rel_\w \kappa}{\mu_\w}
  \frac{\Delta p}{\Delta x}, \\ \frac{Q_\n}{A} &= -
  \frac{\kappa^\rel_\n \kappa}{\mu_\n} \frac{\Delta p}{\Delta x},
\end{align} \end{linenomath} 
where $\kappa$ is the absolute permeability.

Computed relative permeabilities for a subset of the simulations are
plotted in Figure~\ref{fig:kr}, against saturation $S_\w$ and the
non-dimensional pressure gradient $\Pi$. Specifically,
Figure~\ref{fig:kr_w_M_1_pnm}, Figure~\ref{fig:kr_w_M_0_25_pnm} and
Figure~\ref{fig:kr_w_M_4_pnm} show relative permeabilities for the
wetting phase and viscosity ratios $M$ of $1$, $0.25$ and $4$,
respectively. Relative permeabilities for the non-wetting phase and a
viscosity ratio of $4$ is shown in Figure~\ref{fig:kr_n_M_4_pnm}.

In all of these figures, i.e.\ for each value of $M$, the data fall on
a single well-defined surface. This shows that the relative
permeabilities are indeed determined by the three dimensionless
variables $S_\w$, $M$ and $\Pi$, in agreement with the dimensional
analysis in Section~\ref{sec:dimensional_analysis}. \citet{Bardon1980}
mention that gravity (Bond number), wettability (contact angle) and
inertia (Reynolds number) could also affect the relative
permeabilities. These effects are not considered in the simulations
run here, though gravity could be included in the model with relative
ease. 

When measuring relative permeabilities \citep{Oak1990,Bennion2005} and
when using relative permeability models to do continuum-scale
calculations, it is often only their dependence on $S_\w$ which is
considered. It is, however, well-established that variation with $M$
and $\ca$ cannot, in general, be neglected
\citep{Avraam1995,Bardon1980}. Here, the calculated relative
permeabilities are strongly dependent on $\Pi$, and they increase with
increasing $\Pi$. In Figure~\ref{fig:kr}, the color scheme shows that
there is a strong correlation between $\Pi$ and the capillary number,
where high values of $\Pi$ are also associated with high values of
$\ca$. Thus, the results are consistent with those of
\citet{Bardon1980} and \citet{Avraam1995}, who find that relative
permeabilities increase with capillary number. This dependence seems
to disappear, however, as $\Pi \to \infty$. For the viscosity ratios
considered here, the dependence disappears at $\Pi \sim 1$. At this
$\Pi$-value, the average pressure drop over the length $\bar{r}$ is
equal to the typical Young--Laplace interfacial pressure difference,
as discussed in Section~\ref{sec:dimensional_analysis}. The fact that
the relative permeabilities become independent of the pressure
gradient as capillary numbers increase is consistent with the
existence of the high-$\ca$ limit studied by~\citet{Sinha2018}.

According to \citep{Ramstad2012,Bardon1980,Avraam1995}, relative
permeabilities approach straight lines, i.e.\ $\kappa^\rel_\w = S_\w$
and $\kappa^\rel_\n = 1 - S_\w$, at high capillary numbers. In the
equal-viscosity pore network simulations by \citet{Knudsen2002}, this
was found to be the case. Here, however, we find straight lines only
for $M \sim 1$. When $M$ is different from unity, relative
permeabilities converge to non-linear functions of $S_\w$ (and $M$) in
the high-$\ca$ limit.

One of the assumptions in the relative permeability framework is that
the two fluids flow in decoupled flow
channels~\citep{Ramstad2012}. When this is true, it is reasonable that
the permeability of each fluid should be proportional to the cross
sectional area of the porous medium available to it,
i.e.\ proportional to the saturation, when capillary numbers are high.
Such decoupled flow channels are not observed here. Instead, the
fluids exhibit a large degree of mixing at high capillary
numbers. This was observed also by \citet{Sinha2018}, both in pore
network model and lattice-Boltzmann simulations. Disconnected
non-wetting droplets were also observed at high capillary numbers in
the experiments by \citet{Avraam1995}, and were found to contribute
significantly to the total flow rate, although connected pathways were
also present. Our interpretation is therefore that the relative
permeabilities may deviate from straight lines at high capillary
numbers when the fluids mix instead of forming decoupled flow
channels. The effect of this on total mobility and fractional flow is
discussed further in Section~\ref{sec:mobilities} and
Section~\ref{sec:fractional_flows}, respectively.

From Figure~\ref{fig:kr}, it is evident that the relative
permeabilities follow a non-linear curve not unlike those produced by
the classical Corey-type correlations for the lowest capillary
numbers. When working with such correlations, it is typically assumed
that there exists a low-capillary number limit below which relative
permeabilities become independent of flow rate, and the correlations
are valid (for the fluids used in the
measurements). \citet{Ramstad2012} mentions that viscous forces start
to influence the fluid transport at capillary numbers around
$\SI{e-5}{}$. This is consistent with the findings here, which are
that relative permeabilities have a dependence on $\Pi$ down to the
lowest capillary numbers considered of approximately $\SI{e-4}{}$. We
emphasize that the definition of capillary number used here differs
from that used in \citep{Ramstad2012}, since it includes the
porosity. Adoption of the definition from \citep{Ramstad2012} would
reduce all capillary numbers reported here by approximately half an
order of magnitude.

\citet{Avraam1995} find from their experiments that both relative
permeabilities increase with $M$. This is not the case here, at least
not at high capillary numbers. They attribute this to effect to the
existence of films of the wetting fluid, which are not included in our
model.

\subsection{Residual saturations}
\label{sec:residual_saturations}

\begin{figure}[tbp]
  \centering
  \begin{subfigure}[b]{0.48\textwidth}
    \includegraphics[width=\textwidth]{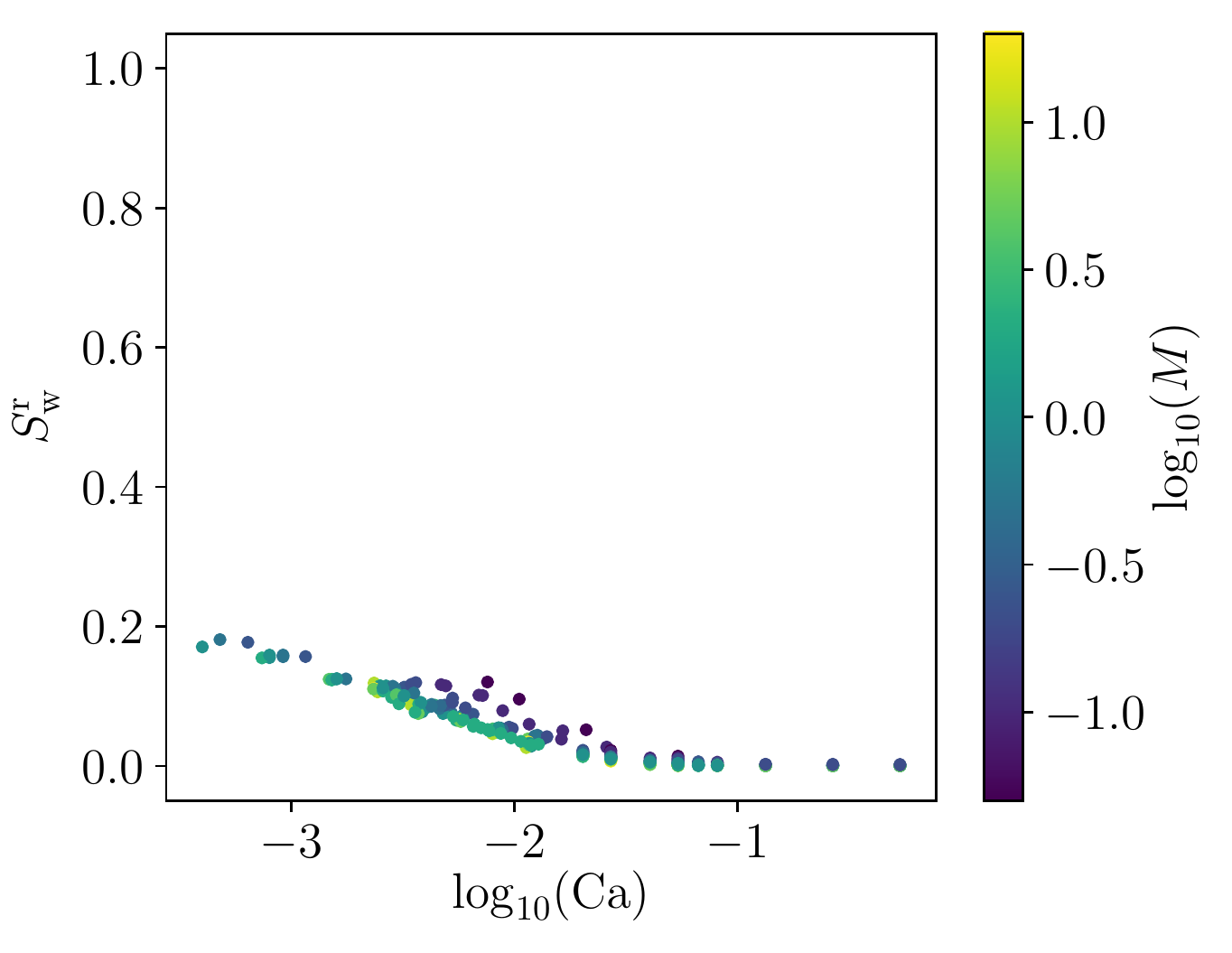}
    \caption{}
  \end{subfigure}
  \begin{subfigure}[b]{0.48\textwidth}
    \includegraphics[width=\textwidth]{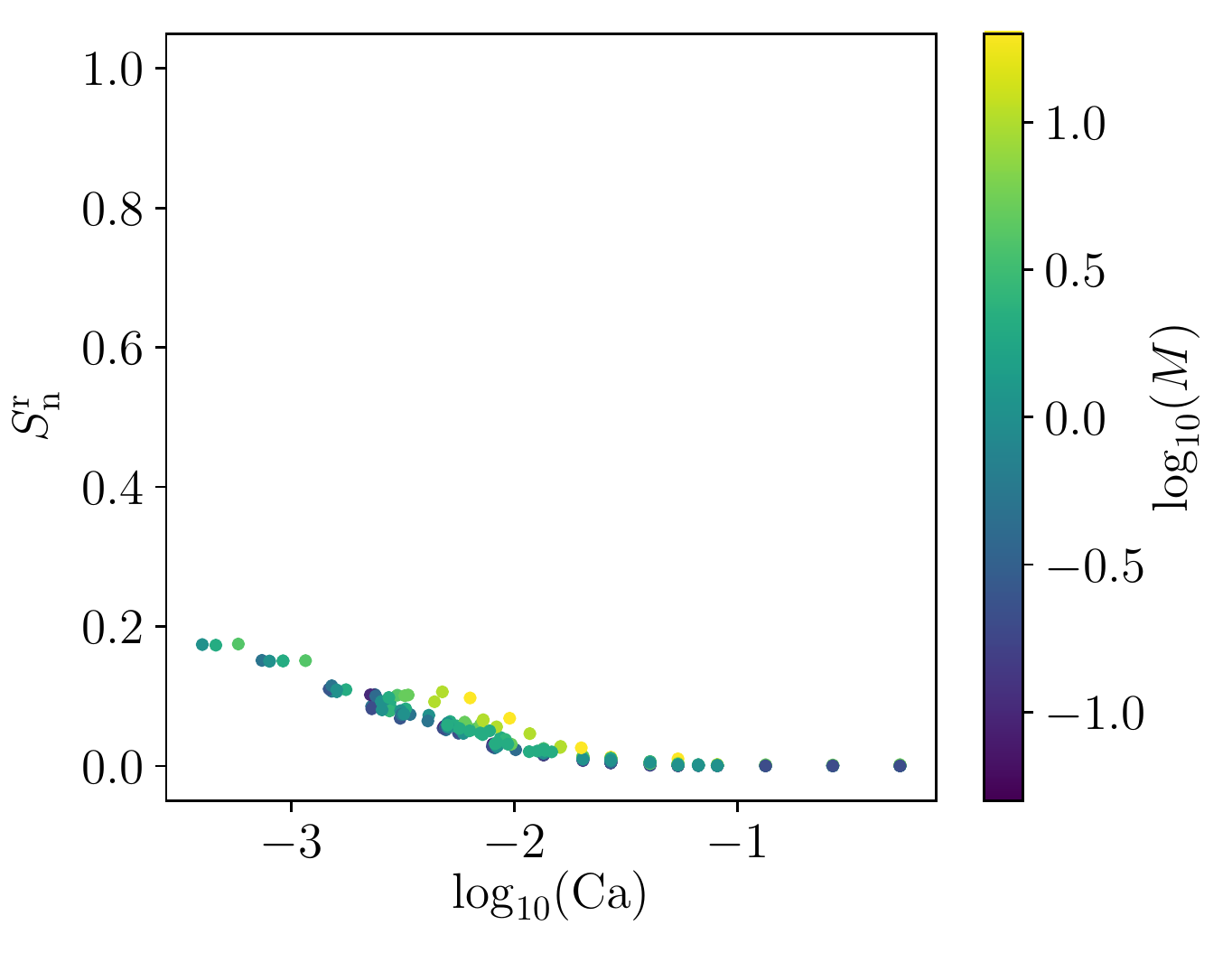}
    \caption{}
  \end{subfigure}
  \caption{Residual saturations for (a) the wetting fluid and (b) the
    non-wetting fluid.}
  \label{fig:S_wr_S_nr}
\end{figure}

For both the wetting and non-wetting fluids, there are regions in
Figure~\ref{fig:kr} where the relative permeabilities are zero. This
was seen by \citet{Knudsen2002} also, for $M=1$. These regions
correspond to irreducible/residual saturations, two other ubiquitous
dimensionless quantities in two-phase porous media flow. The residual
saturations are often defined as the saturation of one fluid that
remains after flooding with the other. This property, defined in this
way, is somewhat difficult to measure using the type of steady-state
pore network model simulations performed here. Therefore, we have
chosen to define the residual saturation of the wetting fluid as the
saturation where the wetting fluid fractional flow falls below
$\SI{e-4}{}$. The residual non-wetting saturation is defined in an
analogous manner. The value of this threshold is somewhat arbitrary,
but it allows for a qualitative discussion.

Computed residual wetting and non-wetting saturations are shown in
Figure~\ref{fig:S_wr_S_nr}. Residual saturations increase as capillary
numbers are reduced, in accordance with findings of
\citet{Ramstad2012} and \citet{Bardon1980}. Furthermore, they reach
zero at at capillary numbers around $0.1$. This means that it is
possible to flush out all of one fluid through flooding with the
other, provided that the flow rate is high enough.

\citet{Bardon1980} also observed that residual saturations were
insensitive to changes in $M$, and this is what we see here
also. Wetting residual saturations are somewhat higher when the
wetting fluid is more viscous and non-wetting residual saturations
are a little higher when the non-wetting fluid is more viscous, but
this effect appears small.

\subsection{Average flow velocity and mobility}
\label{sec:mobilities}

The average mobility $m$ and the average flow velocity $v$ are other
important quantities. They are related through
\begin{linenomath} \begin{align}
  \label{eq:m}
  v &= - m \frac{\Delta p}{\Delta x},
\end{align} \end{linenomath} 
and are discussed together here for reasons that will become apparent
below.

\citet{Sinha2018} studied the high-$\ca$ limit of two-phase porous
media flow. They found that, at high capillary numbers, the average
flow velocity followed a Darcy-type equation,
\begin{linenomath} \begin{align}
  \label{eq:m_D}
  v_\D &= - m_\D \frac{\Delta p}{\Delta x} = -\frac{\kappa}{\bar{\mu}
    \varphi} \frac{\Delta p}{\Delta x},
\end{align} \end{linenomath} 
with an effective viscosity
\begin{linenomath} \begin{align}
  \label{eq:mu_eff}
  \bar{\mu}^\alpha &= S_\w \mu_\w^\alpha + S_\n \mu_\n^\alpha.
\end{align} \end{linenomath} 
The exponent $\alpha$ depended on the degree of mixing of the fluids,
induced by the flow through the porous medium, and was $0.6$ for the
porous medium studied here.

The value $0.6$ of the exponent $\alpha$ is a direct result of the
departure of the relative permeabilities from straight lines at high
capillary numbers. If the relative permeabilities were straight lines,
we would have $\alpha = -1$ and $v_\D/v_0$ could then be expressed as a
linear function of $S_\w$,
\begin{linenomath} \begin{align}
  \left. v_\D/v_0 \right|_{\alpha = -1} &= 1 + S_\w \left( M - 1
  \right),
\end{align} \end{linenomath} 
where $v_0$ is the flow velocity in the single-phase case where $S_\w
= 0$. Instead, with $\alpha = 0.6$, $v_\D/v_0$ is a non-linear
function of $S_\w$.

The existence of this high-$\ca$ limit motivates the study of the
average flow velocity and the average mobility, relative to their limit
values. Dividing \eqref{eq:m} by \eqref{eq:m_D} gives
\begin{linenomath} \begin{align}
  v/v_\D = m/m_\D.
\end{align} \end{linenomath} 
The two quantities $v/v_\D$ and $m/m_\D$ are thus
identical. Moreover, they are dimensionless and may be expected to
vary, roughly, between $0$ and $1$. In particular, they should be $1$
in the two single-phase cases, $S_\w = 1$ and $S_\n = 1$, and in the
high-$\ca$ limit.

\begin{figure}[tbp]
  \centering
  \begin{subfigure}[b]{0.7\textwidth}
    \includegraphics[width=\textwidth]{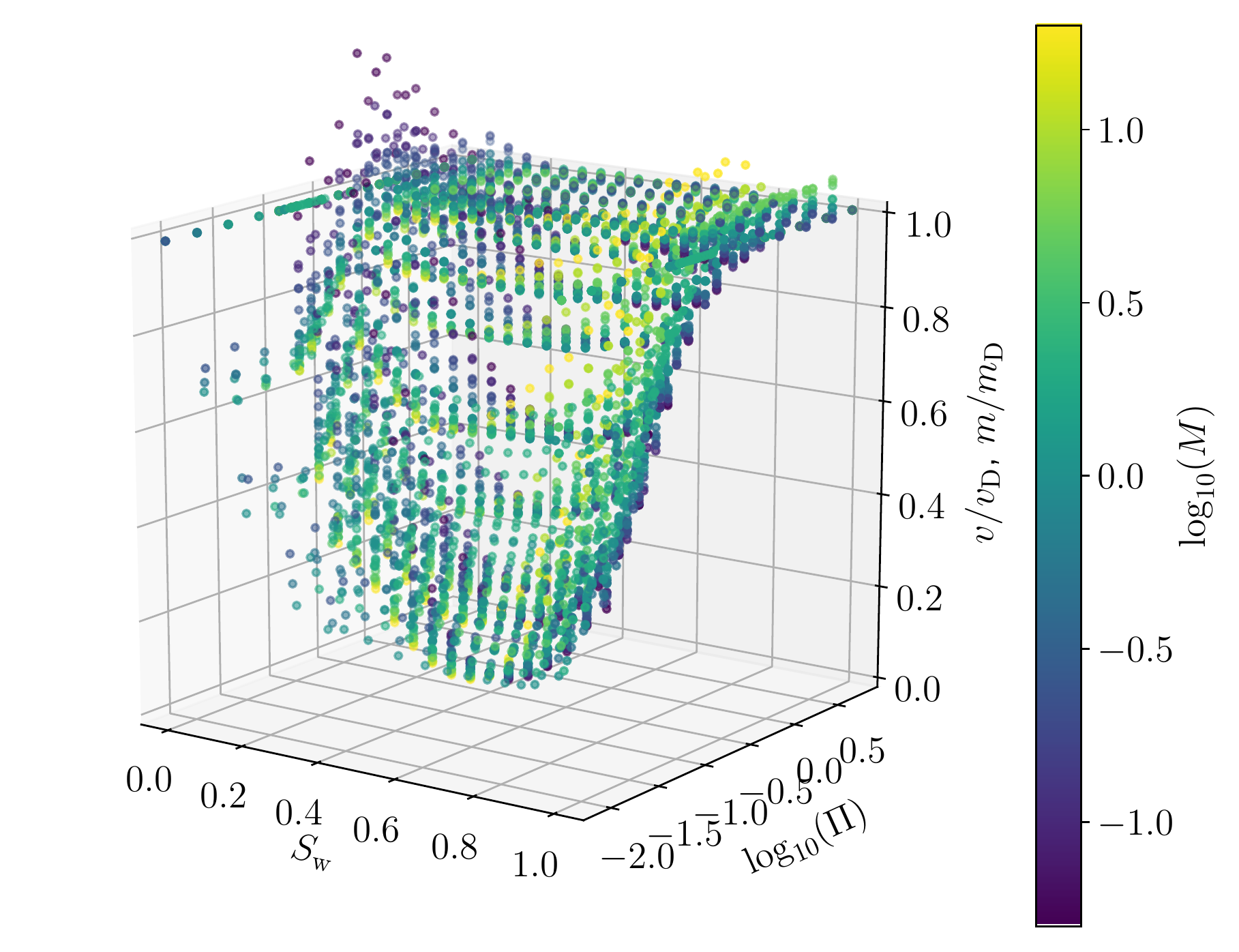}
    \caption{}
    \label{fig:v_v_D}
  \end{subfigure}
  \\
  \begin{subfigure}[b]{0.46\textwidth}
    \includegraphics[width=\textwidth]{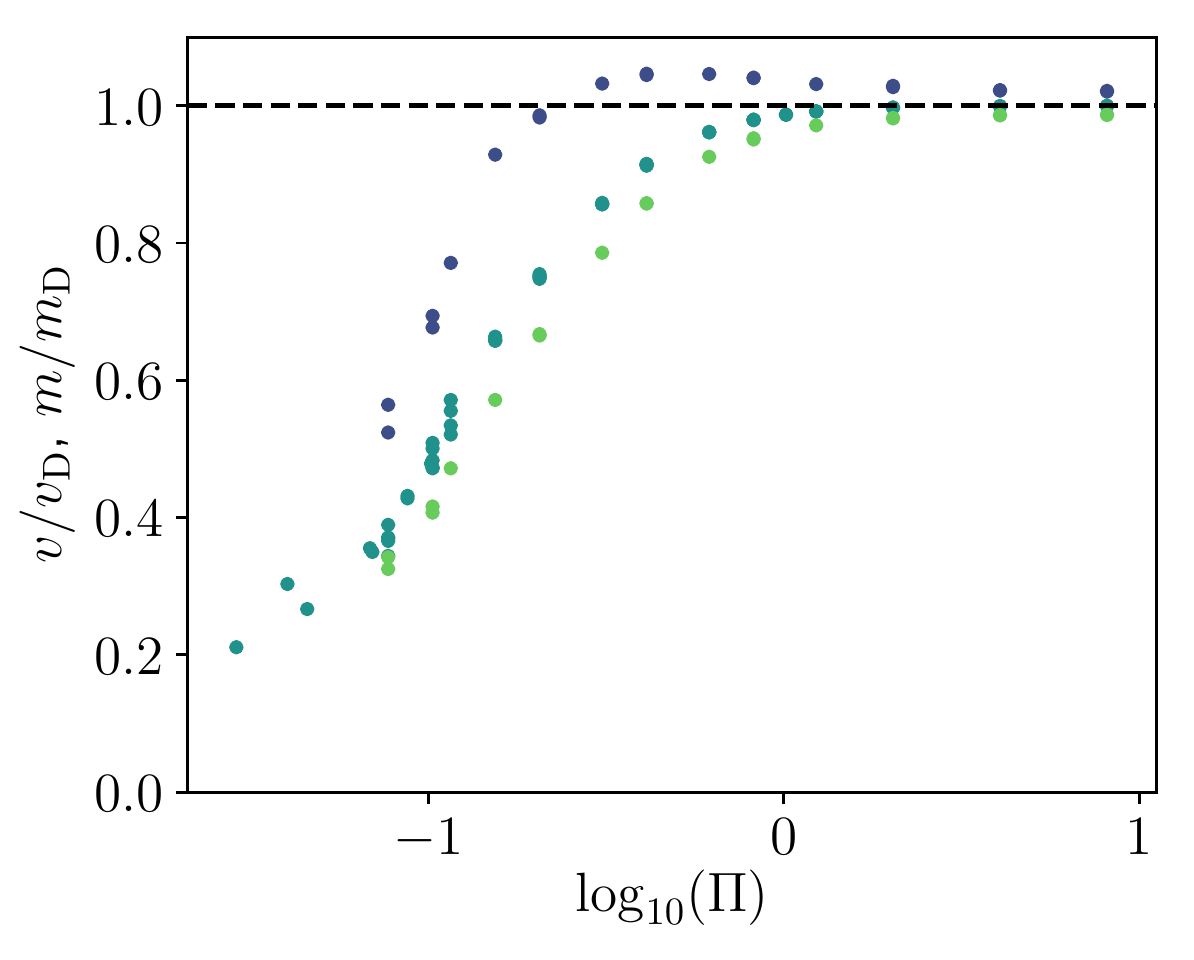}
    \caption{}
    \label{fig:v_v_D_S_w}
  \end{subfigure}
  \begin{subfigure}[b]{0.48\textwidth}
    \includegraphics[width=\textwidth]{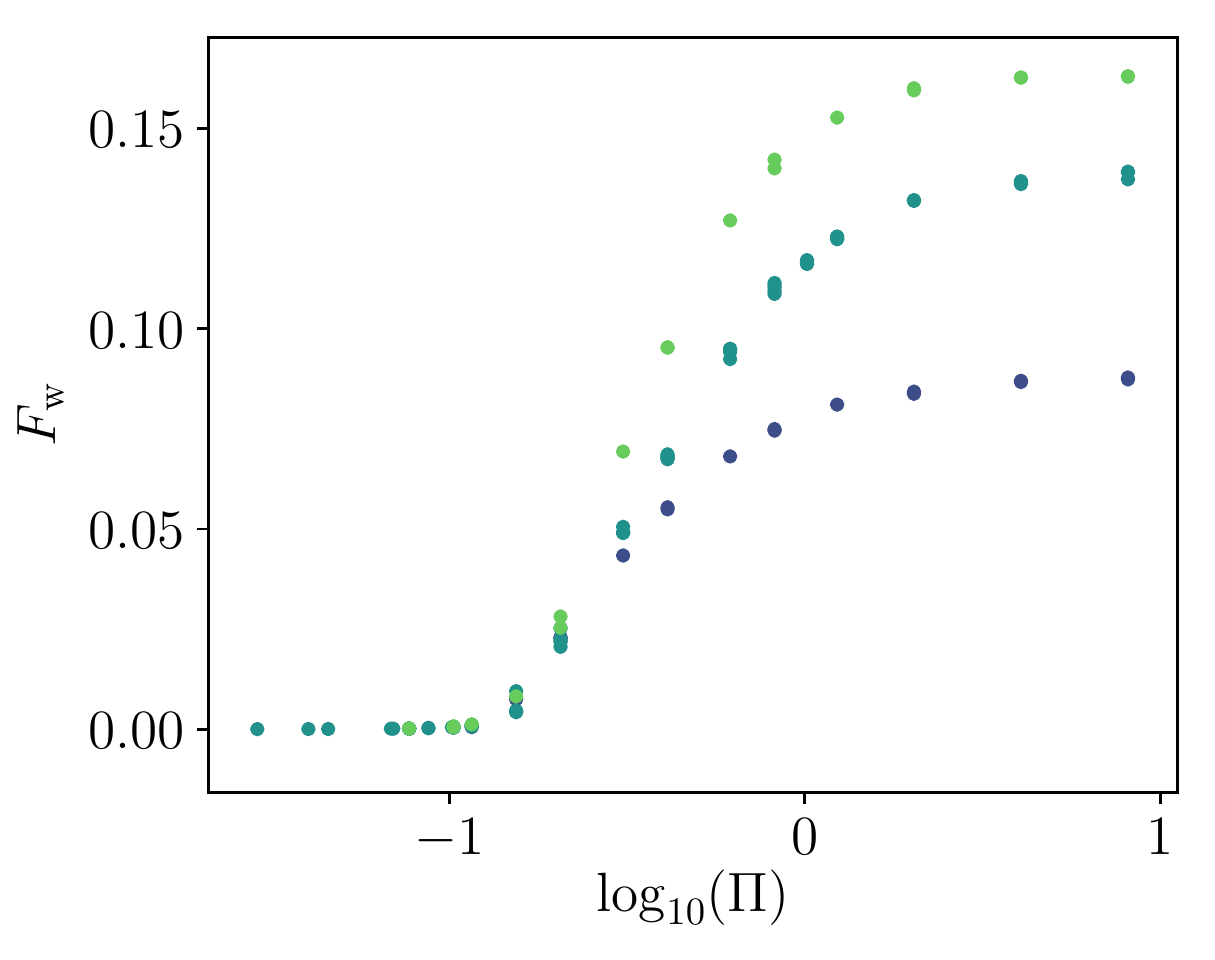}
    \caption{}
    \label{fig:F_w_S_w_0_15}
  \end{subfigure}
  \caption{(a) Calculated values of $v/v_\D$ for all simulations
    performed. (b) Calculated values of $v/v_\D$ for a subset of the
    simulations in (a) with $S_\w = 0.15$ and viscosity ratios of
    $0.2$, $1$ and $5$. (c) Fractional flow for the same set of
    simulations as in (b).}
\end{figure}

Figure~\ref{fig:v_v_D} shows $v/v_\D$ for all simulations run, plotted
against $S_\w$ and $\Pi$. As expected, all data points collapse to $1$
in both single-phase cases. Furthermore, each value of $M$ corresponds
to a single well-defined $v/v_\D$-surface, in accordance with the
dimensional analysis. From the figure, however, it is evident that
these surfaces are not overly sensitive to $M$, at least not for
$S_\w$ around $0.5$. Each constant-$M$ surface reaches values close to
$1$ at the highest values of $\Pi$, in agreement with the findings of
\citet{Sinha2018} for the high-$\ca$ limit.

Interestingly, there are some values of $v/v_\D$ that are larger than
$1$. These occur for the more disparate viscosity ratios, at
saturations where the more viscous fluid is in
minority. Figure~\ref{fig:v_v_D_S_w} shows $v/v_\D$ plotted against
$\Pi$, for $S_\w = 0.15$ and three different viscosity ratios, $0.2$,
$1$ and $5$. The data points with $M=1$ converge to $1$, the limit
value, from below and relatively fast as $\Pi$ increases. The data
points with $M=5$ also approach $1$ from below, but slower. For the
lower viscosity ratio $M=0.2$, on the other hand, $v/v_\D$ increases
fast, overshoots and then approaches $1$ from above.

In Figure~\ref{fig:F_w_S_w_0_15} is shown the fractional flow for the
same data points as in Figure~\ref{fig:v_v_D_S_w}. For the data points
with $M=1$, convergence of $v/v_\D$ to the limit value occurs as the
fractional flow approaches its limit value. The same is true for
$M=0.2$ and $M=0.5$, although convergence is not yet complete for the
largest $\Pi$-values considered.

In terms of mobility ratios $m/m_\D$ these observations may be
understood as follows. At low pressure gradients, all wetting fluid is
stuck, in the sense that $F_\w = 0$, and the non-wetting fluid flows
around it (see Figure~\ref{fig:F_w_S_w_0_15}). As the pressure
gradient is increased, some of the wetting fluid is mobilized and
$F_\w$ increases above zero. This results in more active flow paths
for both fluids and a sharp increase in the average mobility for all
three viscosity ratios.

For $M=0.2$, average mobility reaches a maximum before all wetting
fluid is mobilized, i.e.\ before $F_\w$ converges to its value in the
high-$\ca$ limit. This maximum is caused by the competition between
two different effects. First, an increase in pressure gradient makes
more flow paths available, increasing mobility. Second, $F_\w$
increases and the more viscous wetting fluid makes up a larger
fraction of the flowing fluid. Thus the average viscosity of the
flowing fluid increases, reducing the average mobility. Eventually, a
point is reached where the latter effect becomes more important and a
further increase in the pressure gradient reduces the average
mobility.

For $M=1$, there is no such competition to generate a maximum, as the
wetting and non-wetting fluids are equally viscous and mobilization of
the wetting fluid does not affect the average viscosity.

For $M=5$, the two effects are again present. However, since the
wetting fluid is now less viscous, they both lead to an increase in
mobility with an increase in pressure gradient and we see no
maximum.

The mobility ratios $m/m_\D$ for $M=5$ lie below those for $M=1$ and
they converge slower to the high-$\ca$ limit. A possible reason for
this is that it requires a higher (non-dimensional) pressure gradient
to converge the average viscosity, not changed in the case of $M=1$,
to its high-$\ca$ limit than to mobilize all flow paths in the porous
medium.

\subsection{Fractional flow}
\label{sec:fractional_flows}

\begin{figure}[tbp]
  \centering
  \begin{subfigure}[b]{0.48\textwidth}
    \includegraphics[width=\textwidth]{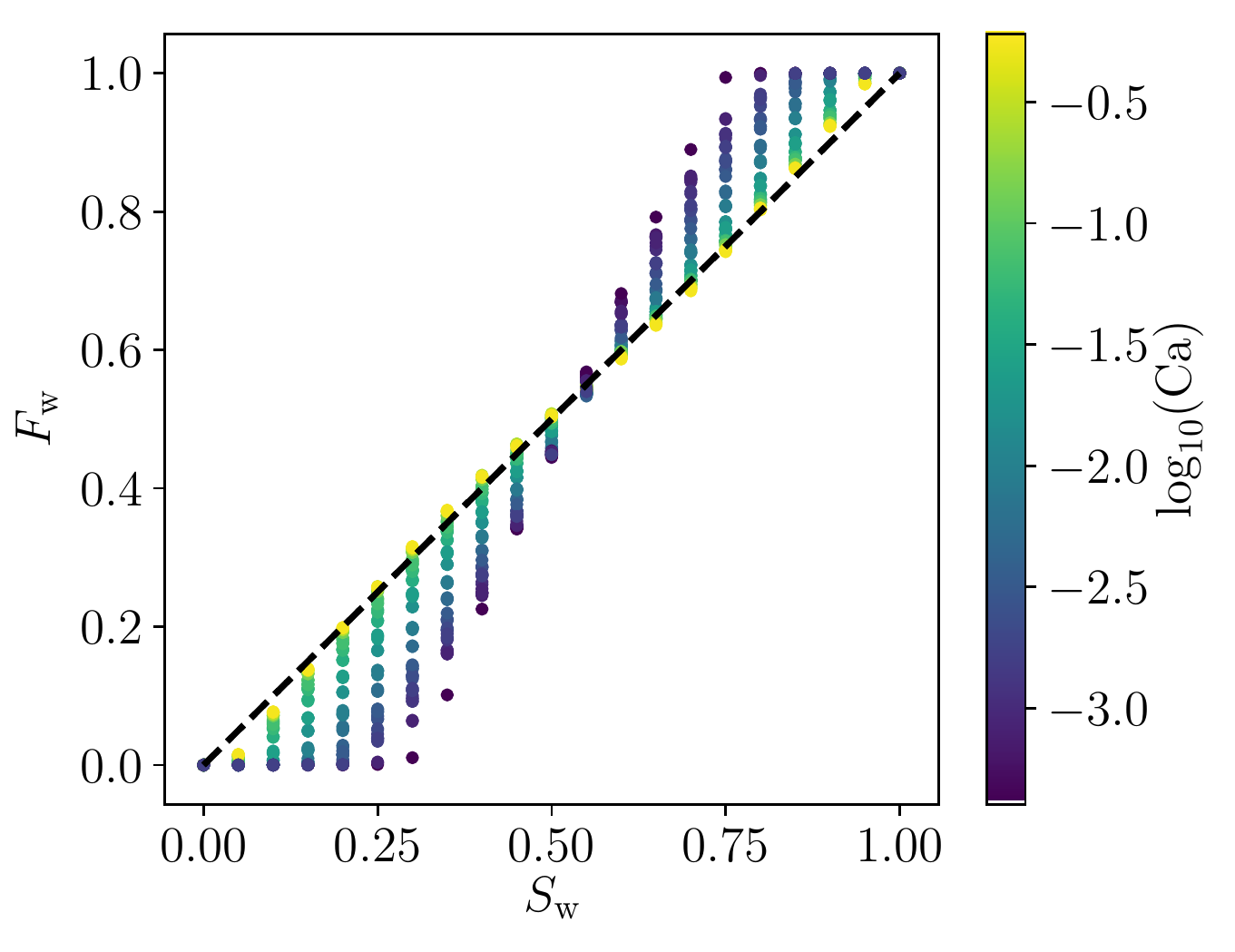}
    \caption{}
    \label{fig:F_w_S_w_M_1}
  \end{subfigure}
  \\
  \begin{subfigure}[b]{0.48\textwidth}
    \includegraphics[width=\textwidth]{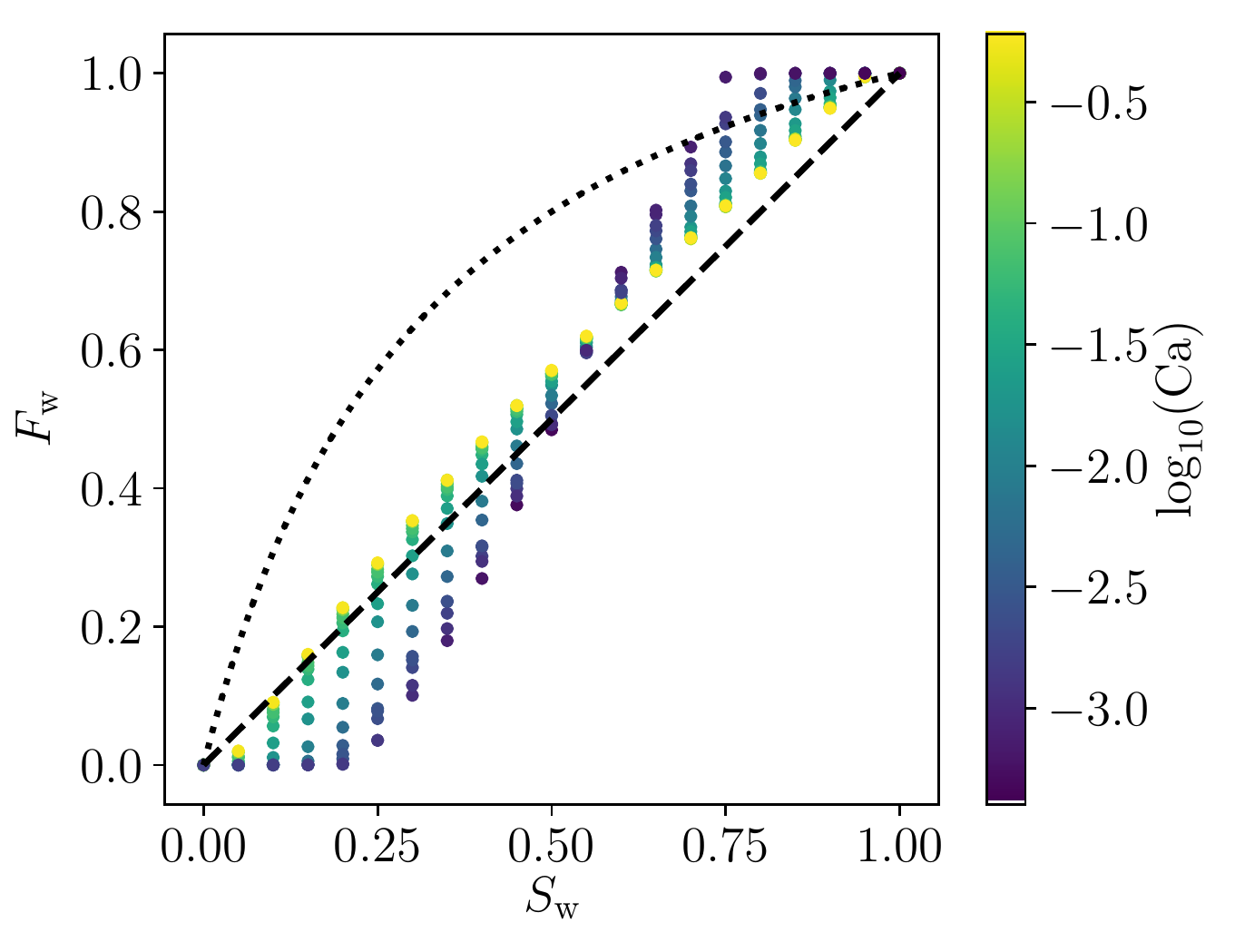}
    \caption{}
    \label{fig:F_w_S_w_M_4}
  \end{subfigure}
  \begin{subfigure}[b]{0.48\textwidth}
    \includegraphics[width=\textwidth]{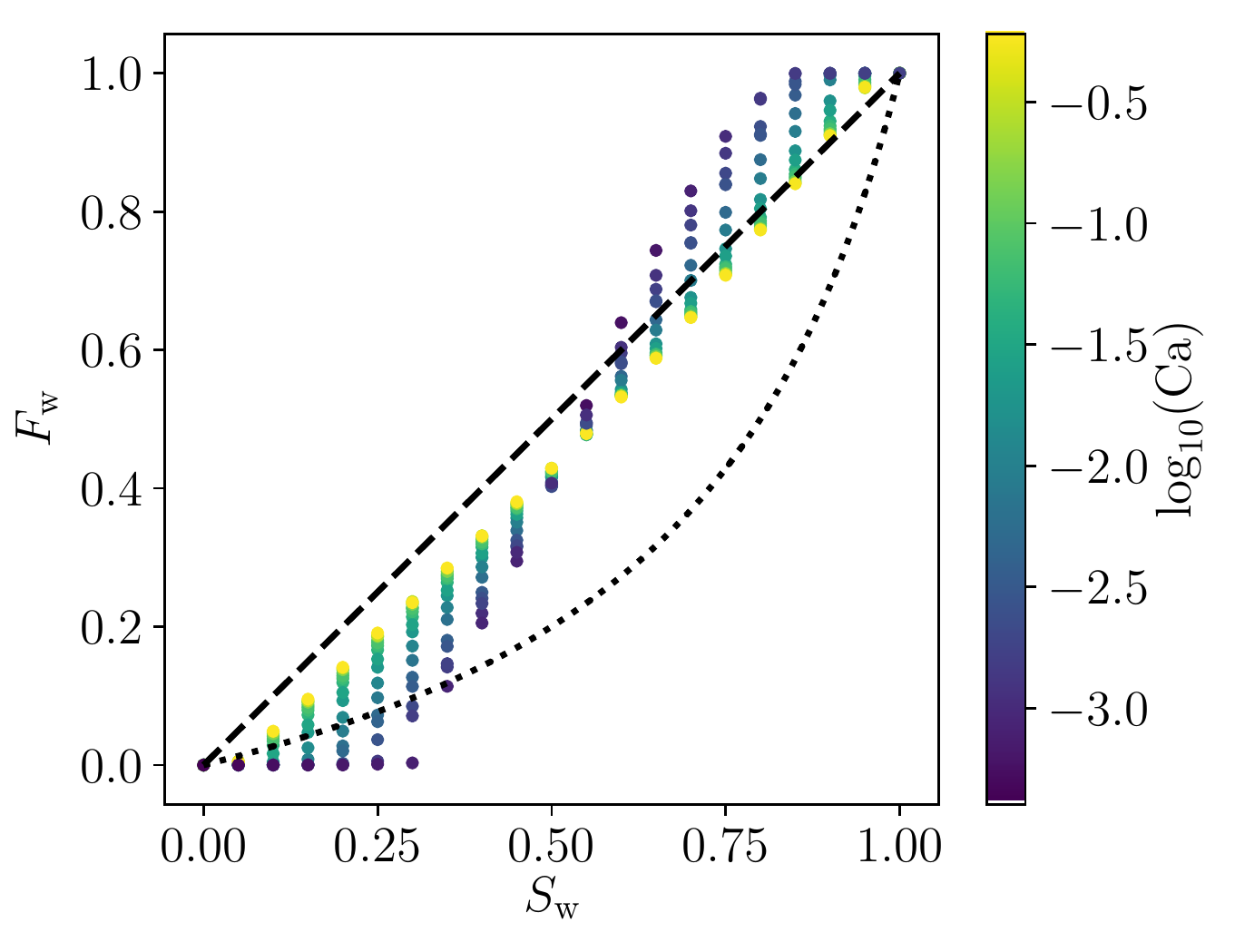}
    \caption{}
    \label{fig:F_w_S_w_M_0_25}
  \end{subfigure}
  \caption{Fractional flow data for (a) $M=1$, (b) $M=4$ and (c)
    $M=0.25$. The dashed lines represent $F_\w = S_\w$ and the dotted
    lines represent the fractional flow obtained if the relative
    permeabilities were $\kappa^\rel_\w = S_\w$ and $\kappa^\rel_\n =
    1 - S_\w$.}
  \label{fig:F_w_S_w}
\end{figure}

The fractional flow for a subset of the performed simulations are
shown in Figure~\ref{fig:F_w_S_w}, for viscosity ratios $1$, $4$, and
$0.25$.

The data in Figure~\ref{fig:F_w_S_w_M_1} for $M=1$ are in qualitative
agreement with those from \citet{Knudsen2002}. They find $F_\w \sim
S_\w$ at high capillary numbers, i.e.\ in the viscosity-dominated
regime. The deviation from the diagonal line representing $F_\w =
S_\w$ increases as the capillary number is reduced. Furthermore,
curves for a specific capillary number are asymmetric w.r.t.\ $S_\w =
0.5$ and cross the diagonal line at $S_\w > 0.5$, meaning that more of
the curve lies below the diagonal than above it. This observation was
explained by \citet{Knudsen2002} by the propensity for the wetting
fluid to occupy narrower pores where flow rate is lower.

By comparing Figure~\ref{fig:F_w_S_w_M_4} and
Figure~\ref{fig:F_w_S_w_M_0_25} we may deduce some of the impact of
the viscosity ratio on the fractional flow. At high capillary numbers,
$F_\w > S_\w$ when $M>1$, i.e.\ when the wetting fluid is less
viscous. Conversely, $F_\w < S_\w$ when $M<1$ and the wetting fluid is
more viscous. The latter was observed also by \citet{Avraam1995}, at
low viscosity ratios and high capillary number, the fractional flow
curves tended to curve upwards.

The dotted lines in Figure~\ref{fig:F_w_S_w_M_4} and
Figure~\ref{fig:F_w_S_w_M_0_25} represent the fractional flows
obtained if the relative permeabilities were $\kappa^\rel_\w = S_\w$
and $\kappa^\rel_\n = 1 - S_\w$, i.e.\ if the fluids followed separate
flow channels. We therefore conclude that mixing of the fluids cause
the flow rates $Q_\w$ and $Q_\n$ to be closer to each other and the
fractional flow curves to lie closer to the diagonal than they would
if the fluids flowed in decoupled flow channels.

At lower capillary numbers, the fractional flow curves obtain the
classical S-shape, as in the case for $M=1$. Also, as is intuitive and
was observed by \citet{Avraam1995}, fractional flow for a given
saturation and capillary number increases with viscosity ratio.

\section{Conclusion}
\label{sec:conclusion}

We have performed more than 6000 steady-state simulations with a
dynamic pore network model of the Aker type \citep{Aker1998},
corresponding to a large span in viscosity ratios and capillary
numbers. From these simulations, dimensionless quantities such as
relative permeabilities, residual saturations, mobility ratios and
fractional flows were computed and discussed. By a dimensional
analysis of the model, all dimensionless output was found to be
functions of the saturation $S_\w$, the viscosity ratio $M$ and the
dimensionless pressure gradient $\Pi$. Effects of wettability, gravity
and inertia were not considered. These effects may add additional
dimensionless variables whose impact could be studied in future work.

Calculated relative permeabilities and residual saturations showed
many of the same qualitative features observed in other experimental
and modeling studies. In particular, the relative permeabilities
increased with capillary numbers and converged to a limit, dependent
on $M$ and $S_\w$, at high capillary numbers. However, while other
studies find that relative permeabilities converge to straight lines
at high capillary numbers we found that this is not the case when $M
\neq 1$. Our conclusion was that departure from straight lines occurs
when fluids mix rather than form decoupled flow channels when
capillary numbers are high. Such mixing behavior has been observed in
previously in pore network and lattice-Boltzmann simulations
\citep{Sinha2018} and, to some extent, in experiments
\citep{Avraam1995}. However, it would be very interesting to see if
experimental studies specifically designed to induce mixing and
measure steady-state properties at high capillary numbers would
produce relative permeability curves that are non-linear in $S_\w$.

Another consequence of the mixing was that computed fractional flow
curves, plotted against saturation, lay closer to the diagonal than
expected from assuming decoupled flow channels. At lower capillary
numbers, fractional flow curves obtained a classical S-shape.

Ratios of average mobility to their high-capillary number limit values
were also considered. These ratios varied, roughly, between 0 and 1,
but values larger than 1 were also observed. For a given saturation
and viscosity ratio, the mobilities were not always monotonically
increasing with the pressure gradient. While increasing the pressure
gradient mobilized more fluid and activates more flow paths, when the
mobilized fluid is more viscous, a reduction in average mobility may
occur instead.

\section*{Acknowledgments}

The authors would like to thank Signe Kjelstrup and Santanu Sinha for
discussions and encouragement. This work was partly supported by the
Research Council of Norway through its Centres of Excellence funding
scheme, project number 262644.

\end{document}